\pdfoutput=1
\documentclass[11pt]{article}

\DeclareUnicodeCharacter{00E9}{\^e}  % é
\DeclareUnicodeCharacter{00E8}{\`e}  % è
\DeclareUnicodeCharacter{00EB}{\"e}  % ë
\DeclareUnicodeCharacter{0113}{\=e}  % ē
\DeclareUnicodeCharacter{0115}{\u{e}}  % ĕ
\DeclareUnicodeCharacter{0117}{\.e}  % ė
\DeclareUnicodeCharacter{0207}{\H{e}}  % ȇ
\DeclareUnicodeCharacter{0205}{\`e}  % ȅ

\DeclareUnicodeCharacter{00EC}{\`i}  % ì
\DeclareUnicodeCharacter{00ED}{\'i}  % í
\DeclareUnicodeCharacter{00EF}{\"i}  % ï
\DeclareUnicodeCharacter{012B}{\=i}  % ī
\DeclareUnicodeCharacter{012D}{\u{i}}  % ĭ
\DeclareUnicodeCharacter{0130}{\.I}  % İ

\DeclareUnicodeCharacter{00F9}{\`u}  % ù
\DeclareUnicodeCharacter{00FC}{\"u}  % ü
\DeclareUnicodeCharacter{016B}{\=u}  % ū
\DeclareUnicodeCharacter{0215}{\u{u}}  % ȕ
\DeclareUnicodeCharacter{0217}{\H{u}}  % ȗ

\DeclareUnicodeCharacter{010D}{\v{c}}  % č
\DeclareUnicodeCharacter{0107}{\'c}  % ć
\DeclareUnicodeCharacter{010B}{\.c}  % ċ
\DeclareUnicodeCharacter{0109}{\^{c}}  % ĉ
\DeclareUnicodeCharacter{1E09}{\d{c}}  % ḉ

\DeclareUnicodeCharacter{015F}{\c{s}}  % ş
\DeclareUnicodeCharacter{015C}{\^{s}}  % ŝ
\DeclareUnicodeCharacter{0161}{\v{s}}  % š
\DeclareUnicodeCharacter{015B}{\'s}  % ś
\DeclareUnicodeCharacter{1E62}{\d{s}}  % ṣ
\DeclareUnicodeCharacter{1E65}{\.{s}}  % ṥ
\DeclareUnicodeCharacter{1E67}{\H{s}}  % ṧ
\DeclareUnicodeCharacter{1E69}{\b{s}}  % ṩ

\DeclareUnicodeCharacter{011F}{g}  % ğ -> g
\DeclareUnicodeCharacter{0131}{i}  % ı -> i
\DeclareUnicodeCharacter{0130}{i}  % İ -> i

\usepackage{ACL2023} %[review]

\usepackage{times}
\usepackage{latexsym}

\usepackage[T1]{fontenc}
\usepackage[utf8]{inputenc}

\usepackage{microtype}
\usepackage[utf8]{inputenc}
\usepackage[T1]{fontenc}
\usepackage{inconsolata}

\usepackage{graphicx}

\title{End-to-End Transformer-based Automatic Speech Recognition for Northern Kurdish: A Pioneering Approach}

\author{Abdulhady Abas Abdullah$^1$ \quad Shima Tabibian$^2$ \quad Hadi Veisi$^3$ \quad Aso Mahmudi$^4$ \quad Tarik Rashid$^1$ \\
	$^1$ University of Kurdistan Hewler \quad $^2$ Cyberspace Research Institute, Shahid Beheshti University \\ $^3$  University of Tehran \quad $^4$ The University of Melbourne\\
	\texttt{abdulhady.abas@ukh.edu.krd, sh}\underline{   }{tabibian@sbu.ac.ir}
 }  

\begin{document}
\maketitle
\begin{abstract}
Automatic Speech Recognition (ASR) for low-resource languages remains a challenging task due to limited training data. This paper introduces a comprehensive study exploring the effectiveness of Whisper, a pre-trained ASR model, for Northern Kurdish (Kurmanji) an under-resourced language spoken in the Middle East. We investigate three fine-tuning strategies: vanilla, specific parameters, and additional modules. Using a Northern Kurdish fin-tuning speech corpus containing approximately 68 hours of validated transcribed data, our experiments demonstrate that the additional module fine-tuning strategy significantly improves ASR accuracy on a specialized test set, achieving a Word Error Rate (WER) of 10.5\% and Character Error Rate (CER) of 5.7\% with Whisper version 3. These results underscore the potential of sophisticated transformer models for low-resource ASR and emphasize the importance of tailored fine-tuning techniques for optimal performance.
\end{abstract}

\section{Introduction}
Automatic speech recognition (ASR) involves converting spoken language into written text. Over the past fifty years, ASR technology has significantly advanced and is now widely applied in various domains, including voice assistants \citep{daneshfar2022pattern}, video transcription, and speech-to-text \citep{zuluaga2023does}. For major international languages like English, ASR systems have achieved near-human-level accuracy, delivering robust, fast, and precise results. However, out of the approximately 7,000 languages spoken worldwide, most lack adequate training resources. Nearly 40\% of these languages are endangered, with fewer than 1,000 speakers each \citep{radford2023robust}. The scarcity of transcribed data for these low-resource languages hampers the training of large neural networks, resulting in poor performance and limited real-world application \cite{abdullah2024advanced}. ASR for low-resource languages has, therefore, become a critical research focus. Recent advancements have introduced specialized models like HuBERT (Hidden-Unit BERT) \citep{hsu2021hubert}, Wav2Vec 2.0 \citep{baevski2020wav2vec}, SeamlessM4T-v2 \citep{barrault2023seamless}, and Whisper \citep{radford2023robust}. These models leverage advanced training techniques to enhance ASR performance in multilingual and low-resource settings. For instance, HuBERT utilizes robust pre-training, Wav2Vec 2.0 employs self-supervised learning for better speech representation, SeamlessM4T-v2 excels in multilingual tasks, and Whisper performs ASR across 100 languages simultaneously, integrating speech translation, language identification, and speech activity detection. 

\begin{figure}
\begin{center}
\includegraphics[scale=0.3]{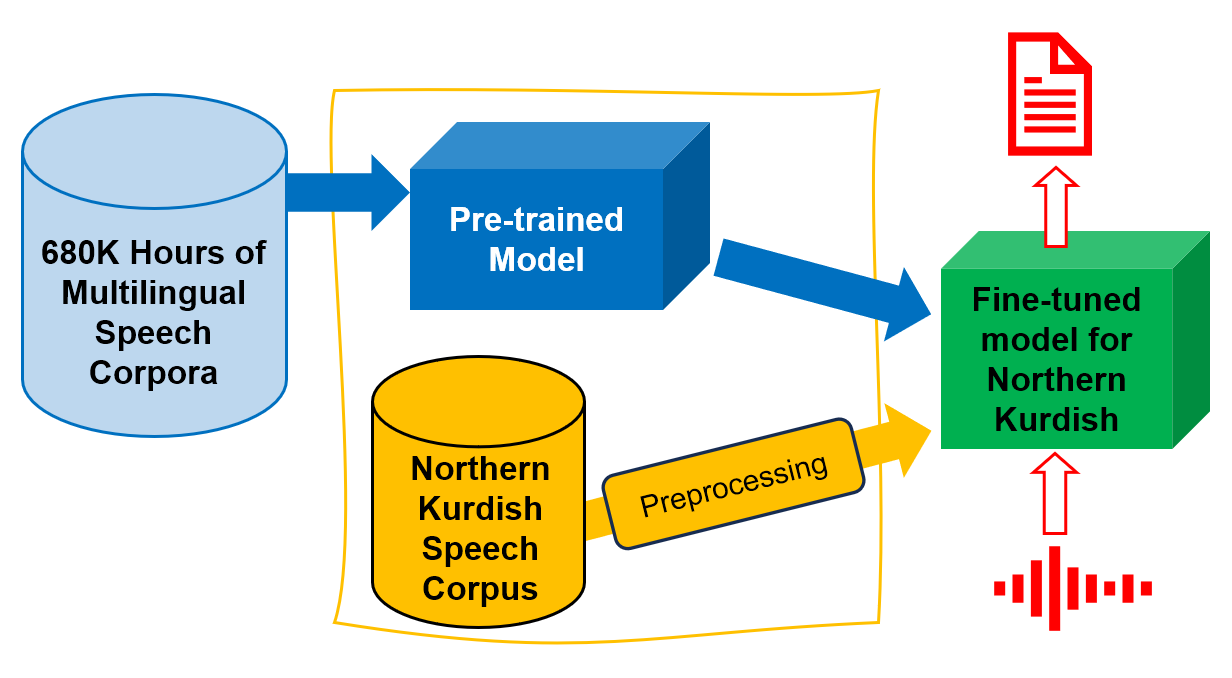} 
\caption{The main architecture of the proposed method.}
\label{fig.model}
\end{center}
\end{figure}

This study investigates the application of Whisper models for ASR in Northern Kurdish (NK), a branch of the Kurdish language, primarily spoken in parts of Turkey, Syria, Iraq, and Iran. The distinct phonetic and grammatical features of NK pose unique challenges and opportunities for ASR systems. We address three key questions in this context:
\begin{enumerate}
 \item To what extent can fine-tuning enhance performance?
 \item Which parts of the model are essential for effective fine-tuning?
 \item What are the pros and cons of parameter-efficient fine-tuning methods?
\end{enumerate}

To explore these questions, we applied three fine-tuning strategies (vanilla, specific parameter, and additional module fine-tuning) on a small, labeled NK dataset. Our results showed substantial performance improvements across all strategies. Vanilla fine-tuning boosted overall performance, specific parameter fine-tuning further enhanced accuracy, and additional module fine-tuning prevented catastrophic forgetting with minimal performance loss, achieving parameter efficiency. The primary contributions of this paper are as follows:

\begin{enumerate}
 \item Exploring Whisper fine-tuning strategies for low-resource ASR
 \item Investigating the internal mechanisms of Whisper's speech encoding
 \item Comparing the advantages and disadvantages of various fine-tuning strategies
 \item Identifying the most effective model for low-resource languages, offering valuable insights for future research and practical applications.
\end{enumerate}

\section{Background}

\subsection{Northern Kurdish}
\begin{figure}
\begin{center}
\includegraphics[scale=0.16]{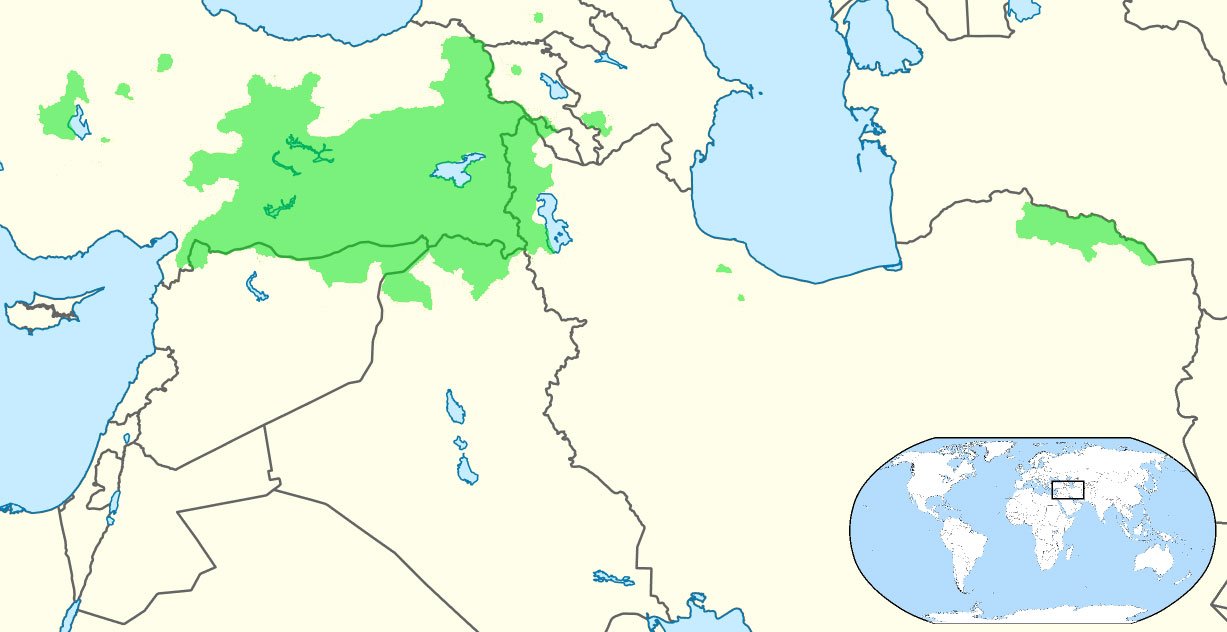} 
\caption{The map of Northern Kurdish-speaking areas (in green) in the Middle East (recreated based on maps in \citet{Murat2023dialects}).}
\label{fig.map}
\end{center}
\end{figure}

Kurdish is a cover term for a cluster of dialects spoken by millions of people in a contiguous area of western Iran, northern Iraq, eastern Turkey, eastern Syria, and separated regions in the Caucasus and Khorasan. 
Among varieties of Kurdish, Northern Kurdish (also known as Kurmanji) and Central Kurdish (also known as Sorani) have more written resources \citep{opengin_2020_3744541}. Although a large population speaks Kurdish, it suffers from the unavailability of sufficient resources for its computational processing purposes. Figure \ref{fig.map} illustrates the geographic distribution of Northern Kurdish in the Middle East region.

Northern Kurdish is written in two alphabets: a modified Latin script and a modified Arabic script. The Latin script is more commonly used, while the Arabic script is mainly used in the Dihok province of Iraq. In this study, we have used the Latin alphabet which includes 31 letters including basic 26 letters and 5 accented letters (ê, î, û, ç, and ş). The current NK alphabet is not entirely phonemic. There are phonemic distinctions in aspirated and unaspirated consonant pairs, such as ç, k, p, and t, as well as between the flap and trill r. Additionally, Arabic loanwords often contain pharyngeal fricatives, which are carefully distinguished in most NK dialects \citep{opengin_2020_3744541}.	

\subsection{Whisper}
OpenAI's Whisper \cite{radford2023robust} is a cutting-edge speech processing model capable of performing tasks like automatic speech recognition (ASR), speech-to-text translation (STT), language identification, and speaker diarization across 100 languages simultaneously. The development team created a weakly supervised dataset of 680,000 hours of speech data, which could be expanded through large-scale collection, automated screening, and processing. They also conducted multi-task standardized annotation on the transcribed text. The team chose a multi-layer stacked Transformer with encoder-decoder structure as the basic network structure.

Depending on the number of layers, the dimension of feature representation (width), and the number of attention heads, the model was divided into five versions (Tiny, Small, Base, Medium, and Large). Table \ref{tab:whisper} summarizes the properties of each version. Large-v2 has 2.5 times more training iterations compared to other large versions. Large-v3 \citep{peng2024owsm} uses data collection and processing like pseudo-labeling with Large-v2 to increase the training data to five million hours. Both large-v2 and large-v3 models outperform the large model, and the large-v3 model demonstrates even stronger capabilities than the large-v2 \citep{zhang2024whisper}. In terms of model training, Whisper uses multi-task training to update and optimize model parameters, including recognition, English translation, speech activation detection, and language identification. Whisper boasts superior multi-language ASR and translation capabilities \citep{wang2023whislu}. In some languages, its performance is even comparable to or better than that of humans. Whisper is becoming increasingly popular due to its advanced features and extensive applications.

\begin{table}
\centering \small
\begin{tabular}{lrrrr}
\textbf{Model} & \textbf{Layers} & \textbf{Width} & \textbf{Heads} & \textbf{Parameters} \\ \hline
Tiny & 4 & 384 & 6 & 39M \\
Base & 6 & 512 & 8 & 74M \\
Small & 12 & 768 & 12 & 244M \\
Medium & 24 & 1024 & 16 & 769M \\
Large & 32 & 1280 & 20 & 1150M \\ \hline
\end{tabular}
\caption{Architecture Details of the Whisper Model Family}
\label{tab:whisper}
\end{table}

\subsection{Fine-tuning}
Fine-tuning involves adjusting the model parameters to fit the hypothesis space of the target task using a much smaller amount of data than the pre-training data. It is typically used when the source and target domains are similar. There are several research studies on using fine-tuning techniques to improve model performance. Well-known self-supervised speech models such as the Wav2Vec series \citep{baevski2020wav2vec}, HuBERT \citep{hsu2021hubert}, and Massively Multilingual Speech (MMS) \citep{barrault2023seamless} require fine-tuning on domain-specific data to adapt to downstream tasks.

However, there are three main challenges with fine-tuning.
Firstly, due to the large parameter size of the models (hundreds of millions or even billions), updating all parameters during fine-tuning can be computationally and time-consuming.
Secondly, fitting a small amount of data with large parameters may lead to overfitting, resulting in poor generalization performance.
Finally, large models have general capabilities for multiple tasks or languages. Still, after fine-tuning, they may only perform well on the target task or language, losing their general abilities known as catastrophic forgetting.

Overfitting and catastrophic forgetting can degrade the generalization ability of a trained model. However, the former is specific to a single language train and test dataset, while the latter concerns multiple languages. Numerous research efforts have been made to use improved fine-tuning strategies to address these issues. Rosin et al. used a partial parameter freezing strategy when adapting an ASR system to Germany and explored the impact of different freezing configurations on system performance \citep{pekarek2023replay}. \citet{pasad2021layer} reinitialized the last 1–3 layers of the Wav2Vec 2.0 model, achieving even better results than the pre-trained model. \citet{kannan2019large} introduced a bottleneck adapter into the model, facilitating full adaptation to specific languages.

\section{Related Works}
ASR is one of the domains that has made remarkable progress, particularly with the rise of deep learning and transformer-based models. Nevertheless, low-resource language is still a problem for languages such as NK since the available data is limited and the languages are diverse. This section provides a background of prior research ASR for Kurdish Dialects.

\subsection{ASR for Kurdish Dialects}
%Specifically, over the last decade, there has been a push to create ASR systems for low-resource languages such as Kurdish. Dealing with limited data and different languages has been a primary concern, and several methods have been suggested from the simple Hidden Markov Models (HMM) to the Deep Learning methods.

\citet{abdullah2022central} introduced a Central Kurdish ASR system based on deep learning techniques. They emphasized the importance of collecting extensive data, using a large text database and a 43-hour AsoSoft speech database to develop reliable ASR models. Their approach combined Recurrent Neural Networks (RNNs) and Convolutional Neural Networks (CNNs) to improve recognition accuracy. Key insights from their study include the critical role of transfer learning and language model adaptation in enhancing ASR performance for low-resource languages like Kurdish.

\citet{veisi2022jira} developed an ASR system named Jira for Central Kurdish, utilizing a Kurdish speech corpus and a pronunciation dictionary that captures the unique characteristics of Kurdish phonemes. Their work explored both traditional Hidden Markov Models (HMMs) and neural network-based models, achieving performance comparable to state-of-the-art ASR systems. This study demonstrated the effectiveness of integrating diverse data sources and modeling approaches to address low-resource language ASR challenges.

Another study, by \citet{gupta2022progress}, focused on multilingual speech recognition for low-resource languages such as Northern Kurdish, Cree, and Inuktitut. Using a Common Voice dataset with approximately 68 hours of Kurmanji speech data, they fine-tuned the Wav2Vec 2.0 model. Their adapted XLSR-53 model achieved a Word Error Rate (WER) of 16\%, highlighting the viability of multilingual approaches for these languages.

Further advancing Kurdish ASR, \citet{abdullah2024breaking} implemented an end-to-end transformer-based Wav2Vec 2.0 model for Central Kurdish. Leveraging a 100-hour speech corpus, they achieved a WER as low as 10\%, marking a significant breakthrough for ASR in this low-resource language.

These studies reflect substantial progress in Kurdish ASR by leveraging deep learning, extensive data collection, and advanced modeling strategies. They underscore the potential to develop effective ASR systems for Kurdish and other low-resource languages, thus advancing the field.

\section{Proposed method}
In this study, an end-to-end transformer-based ASR system has been developed for NK. The system leverages OpenAI’s Whisper model, which is multi-task capable and extensively pre-trained on diverse datasets. To address the challenges of adapting to NK, we fine-tuned Whisper using an NK speech corpus, capturing the phonetic and grammatical nuances unique to the language. This process included character and numerical scaling, as well as fine-tuning specific parameters for enhanced performance. Our approach aimed to improve recognition accuracy and increase the model's resilience to adversarial perturbations, showcasing the capability of transformers in handling low-resource languages. Figure \ref{fig.model} illustrates the core architecture of our method, which includes pre-processing, pre-training, and fine-tuning stages. These stages are further detailed in the subsequent sections.

\subsection{Preprocessing phase}
Text normalization is a crucial step in ASR preprocessing to improve model input. For NK, specific procedures are followed to ensure the text data is consistent and free of variations that could affect ASR performance. This section outlines the steps for standardizing and normalizing NK sentences.

% \paragraph{Character Normalization}
The Latin alphabet for Kurdish includes five accented letters ('ê', 'î', 'û', 'ç', and 'ş'). However, users sometimes type using non-standard keyboards or keyboard layouts from other languages, leading to non-standard forms of these characters. Therefore, it’s important to recognize these variations and replace them with the standard forms. We replaced ['éèëēĕėȇȅ'], ['ìíïīĭİ'], ['ûùüūȕȗ'], ['čćċĉḉ'], and [ŝšśṣṥṧṩ] with 'ê', 'î', 'û', 'ç', and 'ş', respectively.
As most NK users have the Turkish keyboard layout installed on their devices, some Turkish-specific characters appear in NK corpora. We also replaced ['ğ'] with `g', and [İı] with `i'.

\paragraph{Numerical Normalization}
Numerical normalization is a critical preprocessing step for NK text that ensures that numerical data is consistently formatted and easily comprehensible \citep{jamshidi2022hybrid}. This process involves converting integers, floating-point numbers, percentages, and currencies into Kurdish words as they would be spoken aloud. By transforming infinite numeral tokens into a finite set of standard words, this normalization step helps maintain consistency, enhances clarity, and improves ASR application performance by mitigating out-of-vocabulary (OOV) issues. As a result, it reduces interpretation errors and contributes to more reliable and accurate text processing outcomes. Table \ref{tab:num_normal} shows some examples from the numerical normalization step.

\begin{table}
\centering 
\begin{tabular}{p{0.2\linewidth} p{0.6\linewidth}}
\textbf{Original string} & \textbf{Normalized string} \\ \hline
1,234 & \textit{hezar û du sed û sih û çar} \\
12.34 & \textit{duwazdeh poynt sih û çar} \\
\$45.67 & \textit{çil û pênc poynt şêst û heft dolaran} \\
85\% & \textit{heştê û pênc ji sed} \\
-123 & \textit{naqis sed û bîst û sê} \\ \hline 
\end{tabular}
\caption{A few examples from the numerical normalization process.}
\label{tab:num_normal}
\end{table}

\subsection{Pre-training}
\begin{figure}
\begin{center}
\includegraphics[scale=0.25]{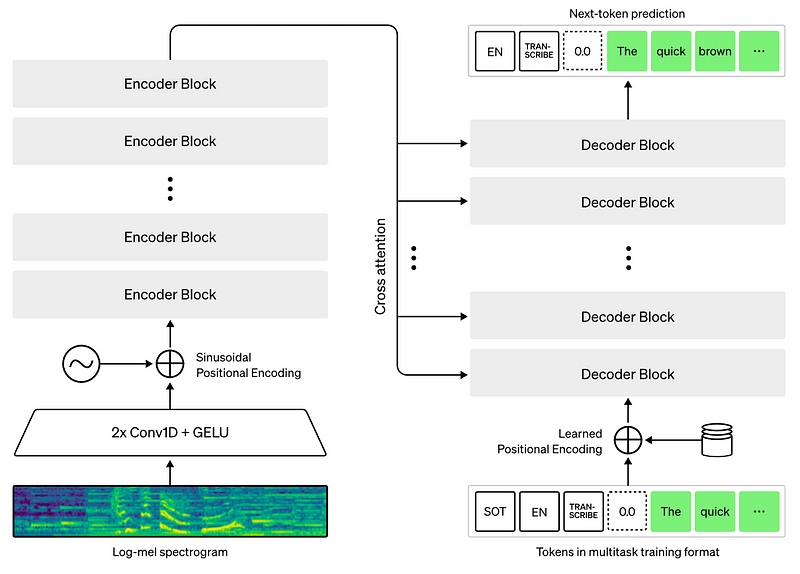} 
\caption{Whisper large -v3 \citep{radford2023robust}}
\label{fig.whisper}
\end{center}
\end{figure}

The Whisper model uses a sequence-to-sequence approach, converting audio spectrogram features into text tokens. This study employed the Whisper large-v3 model—-a Transformer-based encoder-decoder developed by OpenAI, as illustrated in Figure \ref{fig.whisper}. This model is a significant advancement in ASR, trained on over five million hours of labeled and pseudo-labeled data. It improves upon earlier versions by increasing the number of Mel frequency bins in the spectrogram input while maintaining the same core architecture as the large and large-v2 models. Trained for two epochs on a diverse dataset, Whisper large-v3 excels in generalizing across various domains, even in zero-shot settings, The model achieves a 10\% to 20\% reduction in errors compared to its predecessor. In this work, we fine-tuned the pre-trained Whisper large-v3 model on our NK speech corpus, adapting it to NK's unique phonetic and linguistic features.

\subsection{Fine-tuning}
\begin{figure}
\begin{center}
\includegraphics[scale=0.17]{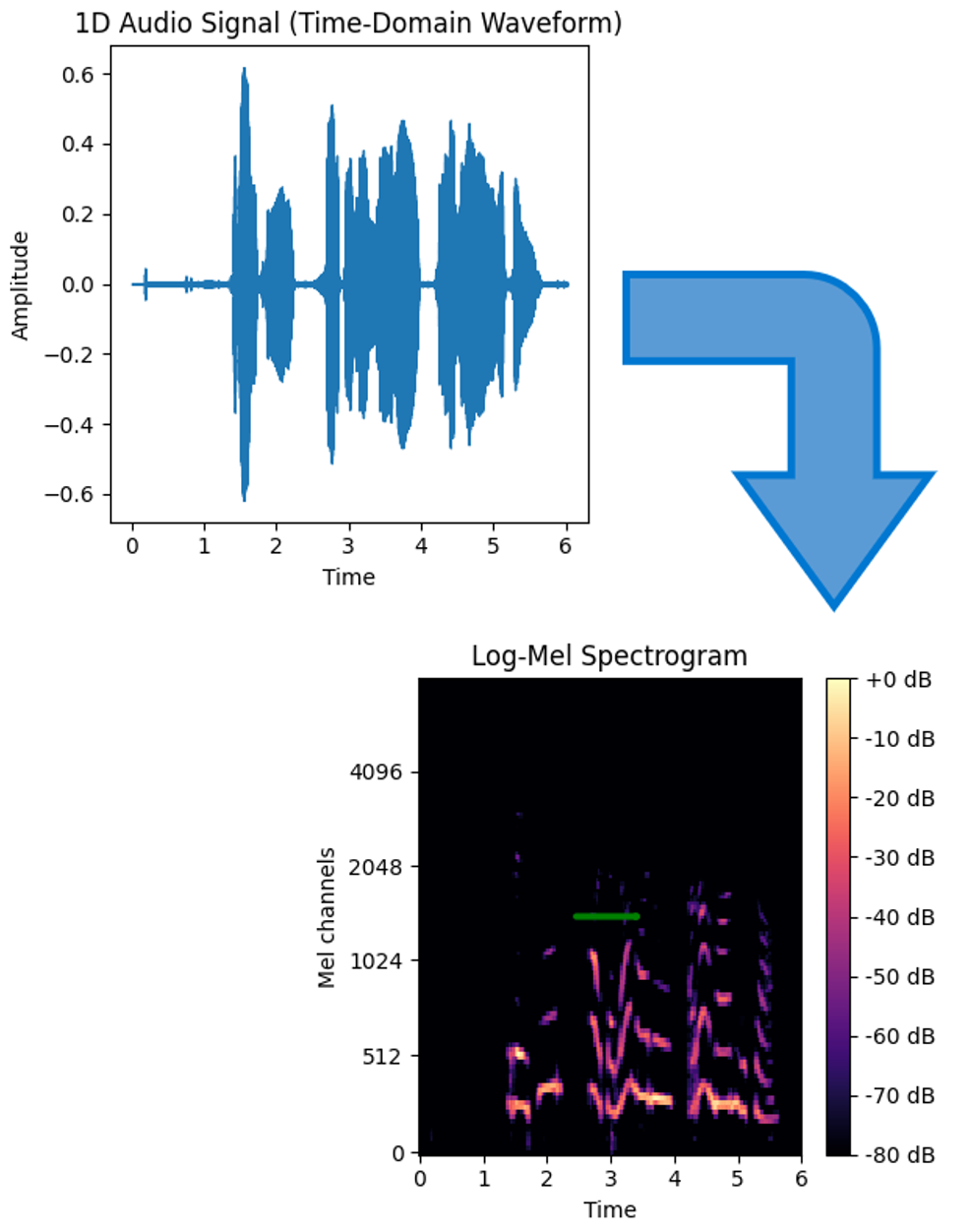} 
\caption{Conversion of sampled speech array to log-Mel spectrogram}
\label{fig.mel}
\end{center}
\end{figure}

We pre-processed the audio data to meet the model's requirements for fine-tuning the Whisper model on our NK speech corpus. This involved converting raw audio into Log-Mel spectrograms (see Figure \ref{fig.mel}), which were fed into the model's encoder. Text transcriptions were tokenized with the Whisper Tokenizer, pre-trained on multiple languages, ensuring broad coverage for our target language.

For fine-tuning, we adjusted the pre-trained Whisper model's parameters using our NK speech corpus, explicitly setting the language and task parameters. The process used a cross-entropy objective, standard for training sequence-to-sequence models. The Trainer handled data collation, gradient accumulation, and metric evaluation over several epochs, assessing performance with the WER metric at each step. Our fine-tuned model showed notable ASR improvements for NK, confirming the effectiveness of the Whisper large-v3 model for low-resource languages. A held-out test set validated the model's ability to generalize well to unseen NK speech. Different fine-tuning methods exploited in this work, are discussed in the following.

\paragraph{1. Vanilla Fine-Tuning}
We began fine-tuning with a Vanilla Fine-Tuning approach, using the entire Whisper model without modifications, allowing all parameters to adapt to the NK speech corpus. The goal was to align the model's pre-trained weights with our dataset. The process involved feeding Log-Mel spectrograms into the encoder and tokenizing transcriptions with the Whisper Tokenizer. Training with a cross-entropy objective function over several epochs helped the model capture NK-specific nuances without added complexities or constraints.

\paragraph{2. Specific Parameter Fine-Tuning}
Building on the vanilla fine-tuning results, we moved to Specific Parameter Fine-Tuning, focusing on most relevant key parameters to our target task, such as attention layers, crucial for contextual understanding in NK. By selectively adjusting these parameters, we reduced overfitting risks while enhancing performance on the NK speech corpus. This targeted approach enabled the model to better capture NK's unique linguistic features.

\paragraph{3. Additional Module Fine-Tuning}
To further boost performance, we implemented Additional Module Fine-Tuning, integrating a newly trained tokenizer based on a 50-million-token NK text corpus. This hybrid tokenizer, combined with the Whisper Tokenizer, improved NK speech tokenization by addressing language-specific challenges. The fine-tuning process involved adjusting both the new tokenizer and additional language modules allowed the model to better capture NK's linguistic nuances, enhancing transcription accuracy and its ability to generalize to unseen data. This underscores the benefit of combining specialized tokenizers with pre-trained model components.

\section{Experiments and analysis}
We conducted experiments to evaluate the performance of various fine-tuning strategies on the Whisper model and to assess our end-to-end transformer-based ASR system for NK. These experiments used the NK Speech Corpus. The following subsections describe the corpus, experiment setups, evaluation metrics, and our findings.

\subsection{Northern Kurdish Speech Corpus}
\paragraph{A) Training and Validation Sets}
The NK Speech Corpus, a key part of the Common Voice Corpus 18.0 \citep{ardila2019common}, plays a vital role in advancing speech technology for NK. Released on June 18, 2024, this comprehensive dataset includes 1.78 GB of audio data, with 101 hours recorded and 68 hours rigorously validated for quality. It is freely available under the CC-0 license, enabling use in both academic and commercial projects. The dataset comprises 567 unique voices, provided in MP3 format to ensure accessibility.

Although nearly half (48\%) of the data lacks specific demographic details, it is evident that the contributions are biased toward younger individuals (aged 20-29) and predominantly male (with 43\% male contributors compared to only 9\% female) \footnote{\href{https://commonvoice.mozilla.org/en/datasets}{https://commonvoice.mozilla.org/en/datasets}}.    

\paragraph{B) Test Set}
The test set was carefully designed to ensure an accurate evaluation of the Kurdish ASR model's performance. It comprises sentences and speakers not included in the training set, ensuring the model is tested on its ability to generalize beyond the training data. We selected 200 diverse sentences for the test set to cover a wide range of linguistic contexts and speech scenarios, offering a broad representation of the language. These 200 sentences were recorded by 50 unique speakers, adding variety in pronunciation, accent, and speaking style. This diversity is crucial for assessing how well the model generalizes across different voices and dialects. By including varied speakers, the test set simulates real-world conditions, providing a realistic benchmark for the model's performance. Full details of the test set are presented in Table \ref{tab:specs}.

\begin{table}
\centering
\begin{tabular}{ll}
\textbf{Specification}    & \textbf{Details} \\ \hline
Total Number of Sentences & 200              \\
Total Number of Speakers  & 50               \\
Duration (Hours)          & 03:25:00         \\
Frequency                 & 22.05 kHz        \\
Sampling Resolution       & 16 Bit, Mono     \\
Format                    & MS WAV      \\ \hline    
\end{tabular}
\caption{The overall specification of the speech test set}
\label{tab:specs}
\end{table}

\begin{table*}
\centering
\begin{tabular}{llll}
\textbf{Model Version} & \textbf{Fine-Tuning  Strategy} & \textbf{WER} & \textbf{CER} \\ \hline\hline
Whisper V1 & Vanilla Fine-Tuning & 14.5\% & 8.2\% \\
Whisper V1 & Specific Parameter Fine-Tuning & 13.1\% & 7.8\% \\
Whisper V1 & Additional Module Fine-Tuning & 12.0\% & 7.0\% \\\hline
Whisper V2 & Vanilla Fine-Tuning & 13.8\% & 7.9\% \\
Whisper V2 & Specific Parameter Fine-Tuning & 12.2\% & 6.7\% \\
Whisper V2 & Additional Module Fine-Tuning & 11.3\% & 6.2\% \\\hline
Whisper V3 & Vanilla Fine-Tuning & 13.2\% & 7.5\% \\
Whisper V3 & Specific Parameter Fine-Tuning & 11.8\% & 6.3\% \\
Whisper V3 & Additional Module Fine-Tuning & \textbf{10.5\%} & \textbf{5.7\%} \\ \hline
\end{tabular}
\caption{The performance comparison of fine-tuning strategies and Whisper model versions}
\label{tab:results}
\end{table*}

\subsection{ASR Evaluation Criteria}
ASR systems are evaluated using several key metrics to determine their effectiveness, accuracy, and suitability for specific applications. The following are basic assessment criteria commonly used in ASR research and development:

\paragraph{Word Error Rate (WER)} is the most widely used metric for ASR systems. It calculates the proportion of words incorrectly recognized by the ASR system compared to a reference transcript. A lower WER indicates higher recognition accuracy, providing a clear view of the errors in the ASR output relative to all spoken words.

\paragraph{Character Error Rate (CER)} assesses ASR accuracy at the character level, calculating the percentage of characters misrecognized. CER is particularly useful for logographic languages and applications where character-level accuracy is crucial, such as handwriting recognition. This metric is beneficial for cases where precise character recognition significantly impacts system performance.

WER and CER are calculated as:
\[\frac{S+D+I}{N} * 100\]
where \(S\), \(D\), \(I\), and \(N\) represent the number of substitutions, deletions, insertions, and total words/characters in the reference transcript, respectively. 

\subsection{Experimental Setup}
The NK ASR corpus includes about 68 hours of audio, with 90\% used for training to expose the model to diverse voice samples. The remaining 10\% is set aside for evaluation, monitoring performance, and tuning hyper-parameters. A controlled set of 200 sentences, recorded by 50 native speakers, introduces variation in pronunciation, accent, and style to prevent overfitting and ensure a fair assessment of the model’s generalizability.

Training utilized two NVIDIA RTX 4090 GPUs in SLI mode, offering enhanced processing power and memory bandwidth. The system’s 290 GB RAM accommodated large data and models.

DeepSpeed was employed to maximize RAM and GPU efficiency during training. By leveraging features like ZeRO, which distributes model states across GPUs, DeepSpeed minimizes memory redundancy, allowing for larger models and batch sizes. Gradient checkpointing and mixed precision were also implemented, which reduce memory usage by recomputing certain activations as needed and performing lower-precision calculations. These optimizations prevent memory overload, accelerate training by using multiple GPUs in parallel, shorten epoch times, and increase convergence rates by efficiently handling large datasets.

Given the complex training process—which involves multiple encoding rounds and dataset partitioning for training, validation, and testing—the model’s structure required careful optimization. As a result, training spanned several epochs, with critical metrics such as loss and accuracy monitored throughout.

Hyper-parameter tuning was essential, with settings including a 1e-5 base learning rate, 500 warmup steps, and a maximum of 500,000 steps to ensure stable and thorough training. Fine-tuning involved:
\begin{enumerate}
    \item \textbf{Preparation}: Ensure Whisper model and NK data are preprocessed and ready.
    \item \textbf{Configuration}: Adjust the training script with DeepSpeed optimizations, a learning rate of 1e-5, 500 warmup steps, and a 500,000-step maximum.
    \item \textbf{Training}: Use DeepSpeed for efficient distributed processing. Track metrics like loss and accuracy to monitor convergence.
    \item \textbf{Evaluation and Tuning}: Regularly assess the model on validation and test sets, adjusting hyper-parameters as needed.
    \item \textbf{Finalization}: Save the trained model once satisfactory performance is achieved for deployment or further evaluation.
\end{enumerate}
 
\subsection{Result and Discussion}
The performance of the Whisper model fine-tuned with different strategies, was evaluated using the NK Speech Corpus test set. We also compared the results across different versions of the Whisper model pre-trained on English data. The results are summarized in Table \ref{tab:results}.

As shown in Table \ref{tab:results}, the results indicate varying performance levels across different versions of the Whisper model. For Whisper V1, all fine-tuning strategies enhanced performance compared to the baseline but did not achieve the levels reached by later versions. Whisper V2 showed improved results, with additional module fine-tuning yielding a Word Error Rate (WER) of 11.3\% and a Character Error Rate (CER) of 6.2\%. The latest version, Whisper V3, achieved the lowest WER of 10.5\% and a CER of 5.7\% with additional module fine-tuning, demonstrating its superior ability to handle NK phonetics and nuances. Overall, the comparison suggests that while all Whisper model versions benefited from fine-tuning, Whisper V3 with additional module fine-tuning delivered the most significant improvements. This indicates that newer versions of the Whisper model, combined with advanced fine-tuning techniques, provide substantial advantages in processing low-resource languages like NK.

Table \ref{tab:compare-results} compares the performance of two Benchmarks in Northern Kurdish (Kurmanji) ASR tasks: Wav2Vec2.0 XLSR-53 (reported by \citep{gupta2022progress}) and our Whisper V3 model. The Wav2Vec2.0 XLSR-53 achieved a WER of 16.3\% and a CER of 9.2\%, indicating some limitations in accurately encoding all linguistic and phonetic information in NK speech. In contrast, the Whisper V3 Benchmark outperformed it by a significant margin, achieving a 5.8\% lower WER and 3.5\% lower CER. 

\begin{table}
\centering
\begin{tabular}{cll}
\textbf{Model} & \textbf{WER} & \textbf{CER} \\ \hline
Wav2Vec2.0 XLSR-53   & 16.3\% & 9.2\% \\
Whisper V3 Benchmark (ours) & 10.5\% & 5.7\% \\
\hline
\end{tabular}
\caption{Comparison of performance between two Northern Kurdish ASR systems: Wav2Vec2.0 XLSR-53 \citep{gupta2022progress} and our Whisper V3.}
\label{tab:compare-results}
\end{table}

These results indicate that Whisper V3 transcribes NK with significantly lower word and character error rates. Lower-resource languages like NK are handled more effectively than with Whisper V1 and previous methods, thanks to advancements in Whisper V3's architecture and fine-tuning techniques. This establishes a new benchmark, with Whisper V3 outperforming its predecessor, Wav2Vec 2 XLSR-53. Consequently, Whisper V3 is a superior choice for spoken language documentation tasks in low-resource languages due to its enhanced accuracy.

\section{Conclusion}

This study demonstrates the effectiveness of fine-tuning Whisper, a state-of-the-art ASR model, for low-resource languages like Northern Kuddish (Kurmanji). Our findings reveal that fine-tuning significantly improves Whisper's performance, especially in newer versions like Whisper V3. Compared to traditional models, Whisper V3 offers higher accuracy and sensitivity for NK speech. This research highlights the potential of advanced transformer models for addressing language technology gaps in underrepresented languages.

Future work could explore the integration of additional linguistic features and a more diverse dataset to further enhance performance. This study serves as a step toward more accessible ASR technology, with broader implications for language preservation and cultural engagement.

\bibliography{refs}

\begin{thebibliography}{21}
\expandafter\ifx\csname natexlab\endcsname\relax\def\natexlab#1{#1}\fi

\bibitem[{Abdullah et~al.(2024{\natexlab{a}})Abdullah, Ahmed, Rashid, Veisi, Rassul, Hassan, Fattah, Ali, and Shamsaldin}]{abdullah2024advanced}
Abdulhady~Abas Abdullah, Aram~Mahmood Ahmed, Tarik Rashid, Hadi Veisi, Yassin~Hussein Rassul, Bryar Hassan, Polla Fattah, Sabat~Abdulhameed Ali, and Ahmed~S Shamsaldin. 2024{\natexlab{a}}.
\newblock Advanced clustering techniques for speech signal enhancement: A review and metanalysis of fuzzy c-means, k-means, and kernel fuzzy c-means methods.
\newblock \emph{arXiv preprint arXiv:2409.19448}.

\bibitem[{Abdullah and Veisi(2022)}]{abdullah2022central}
Abdulhady~Abas Abdullah and Hadi Veisi. 2022.
\newblock Central kurdish automatic speech recognition using deep learning.
\newblock \emph{Journal of University of Anbar for Pure Science}, 16(2).

\bibitem[{Abdullah et~al.(2024{\natexlab{b}})Abdullah, Veisi, and Rashid}]{abdullah2024breaking}
Abdulhady~Abas Abdullah, Hadi Veisi, and Tarik Rashid. 2024{\natexlab{b}}.
\newblock Breaking walls: Pioneering automatic speech recognition for central kurdish: End-to-end transformer paradigm.
\newblock \emph{arXiv preprint arXiv:2406.02561}.

\bibitem[{Ardila et~al.(2020)Ardila, Branson, Davis, Kohler, Meyer, Henretty, Morais, Saunders, Tyers, and Weber}]{ardila2019common}
Rosana Ardila, Megan Branson, Kelly Davis, Michael Kohler, Josh Meyer, Michael Henretty, Reuben Morais, Lindsay Saunders, Francis Tyers, and Gregor Weber. 2020.
\newblock \href {https://aclanthology.org/2020.lrec-1.520} {Common voice: A massively-multilingual speech corpus}.
\newblock In \emph{Proceedings of the Twelfth Language Resources and Evaluation Conference}, pages 4218--4222, Marseille, France. European Language Resources Association.

\bibitem[{Baevski et~al.(2020)Baevski, Zhou, Mohamed, and Auli}]{baevski2020wav2vec}
Alexei Baevski, Yuhao Zhou, Abdelrahman Mohamed, and Michael Auli. 2020.
\newblock Wav2vec 2.0: A framework for self-supervised learning of speech representations.
\newblock \emph{Advances in neural information processing systems}, 33:12449--12460.

\bibitem[{Baran(2023)}]{Murat2023dialects}
Murat Baran. 2023.
\newblock \href {https://serkeftin.com/en/zaravayen-kurmanci-dialects-of-kurmanji/} {Dialects of kurmanji}.
\newblock Accessed: 2024-10-13.

\bibitem[{Barrault et~al.(2023)Barrault, Chung, Meglioli, Dale, Dong, Duppenthaler, Duquenne, Ellis, Elsahar, Haaheim et~al.}]{barrault2023seamless}
Lo{\"\i}c Barrault, Yu-An Chung, Mariano~Coria Meglioli, David Dale, Ning Dong, Mark Duppenthaler, Paul-Ambroise Duquenne, Brian Ellis, Hady Elsahar, Justin Haaheim, et~al. 2023.
\newblock Seamless: Multilingual expressive and streaming speech translation.
\newblock \emph{arXiv preprint arXiv:2312.05187}.

\bibitem[{Daneshfar and Jamshidi(2022)}]{daneshfar2022pattern}
Fatemeh Daneshfar and Mohammad~Behdad Jamshidi. 2022.
\newblock A pattern recognition framework for signal processing in metaverse.
\newblock In \emph{2022 8th Iranian Conference on Signal Processing and Intelligent Systems (ICSPIS)}, pages 1--5. IEEE.

\bibitem[{Gupta and Boulianne(2022)}]{gupta2022progress}
Vishwa Gupta and Gilles Boulianne. 2022.
\newblock Progress in multilingual speech recognition for low resource languages kurmanji kurdish, cree and inuktut.
\newblock In \emph{Proceedings of the Thirteenth Language Resources and Evaluation Conference}, pages 6420--6428.

\bibitem[{Hsu et~al.(2021)Hsu, Bolte, Tsai, Lakhotia, Salakhutdinov, and Mohamed}]{hsu2021hubert}
Wei-Ning Hsu, Benjamin Bolte, Yao-Hung~Hubert Tsai, Kushal Lakhotia, Ruslan Salakhutdinov, and Abdelrahman Mohamed. 2021.
\newblock Hubert: Self-supervised speech representation learning by masked prediction of hidden units.
\newblock \emph{IEEE/ACM transactions on audio, speech, and language processing}, 29:3451--3460.

\bibitem[{Jamshidi and Daneshfar(2022)}]{jamshidi2022hybrid}
Mohammad~Behdad Jamshidi and Fatemeh Daneshfar. 2022.
\newblock A hybrid echo state network for hypercomplex pattern recognition, classification, and big data analysis.
\newblock In \emph{2022 12th International Conference on Computer and Knowledge Engineering (ICCKE)}, pages 007--012. IEEE.

\bibitem[{Kannan et~al.(2019)Kannan, Datta, Sainath, Weinstein, Ramabhadran, Wu, Bapna, Chen, and Lee}]{kannan2019large}
Anjuli Kannan, Arindrima Datta, Tara~N Sainath, Eugene Weinstein, Bhuvana Ramabhadran, Yonghui Wu, Ankur Bapna, Zhifeng Chen, and Seungji Lee. 2019.
\newblock Large-scale multilingual speech recognition with a streaming end-to-end model.
\newblock \emph{arXiv preprint arXiv:1909.05330}.

\bibitem[{Pasad et~al.(2021)Pasad, Chou, and Livescu}]{pasad2021layer}
Ankita Pasad, Ju-Chieh Chou, and Karen Livescu. 2021.
\newblock Layer-wise analysis of a self-supervised speech representation model.
\newblock In \emph{2021 IEEE Automatic Speech Recognition and Understanding Workshop (ASRU)}, pages 914--921. IEEE.

\bibitem[{Pekarek~Rosin and Wermter(2023)}]{pekarek2023replay}
Theresa Pekarek~Rosin and Stefan Wermter. 2023.
\newblock Replay to remember: Continual layer-specific fine-tuning for german speech recognition.
\newblock In \emph{International Conference on Artificial Neural Networks}, pages 489--500. Springer.

\bibitem[{Peng et~al.(2024)Peng, Tian, Chen, Arora, Yan, Sudo, Shakeel, Choi, Shi, Chang et~al.}]{peng2024owsm}
Yifan Peng, Jinchuan Tian, William Chen, Siddhant Arora, Brian Yan, Yui Sudo, Muhammad Shakeel, Kwanghee Choi, Jiatong Shi, Xuankai Chang, et~al. 2024.
\newblock Owsm v3. 1: Better and faster open whisper-style speech models based on e-branchformer.
\newblock \emph{arXiv preprint arXiv:2401.16658}.

\bibitem[{Radford et~al.(2023)Radford, Kim, Xu, Brockman, McLeavey, and Sutskever}]{radford2023robust}
Alec Radford, Jong~Wook Kim, Tao Xu, Greg Brockman, Christine McLeavey, and Ilya Sutskever. 2023.
\newblock Robust speech recognition via large-scale weak supervision.
\newblock In \emph{International conference on machine learning}, pages 28492--28518. PMLR.

\bibitem[{Veisi et~al.(2022)Veisi, Hosseini, MohammadAmini, Fathy, and Mahmudi}]{veisi2022jira}
Hadi Veisi, Hawre Hosseini, Mohammad MohammadAmini, Wirya Fathy, and Aso Mahmudi. 2022.
\newblock Jira: a central kurdish speech recognition system, designing and building speech corpus and pronunciation lexicon.
\newblock \emph{Language Resources and Evaluation}, 56(3):917--941.

\bibitem[{Wang et~al.(2023)Wang, Li, Guo, Qiao, Li, Shang, Wei, Tao, Zhang, and Yang}]{wang2023whislu}
Minghan Wang, Yinglu Li, Jiaxin Guo, Xiaosong Qiao, Zongyao Li, Hengchao Shang, Daimeng Wei, Shimin Tao, Min Zhang, and Hao Yang. 2023.
\newblock Whislu: End-to-end spoken language understanding with whisper.
\newblock In \emph{Proc. Interspeech}, volume 2023, pages 770--774.

\bibitem[{Zhang et~al.(2024)Zhang, Jiang, Wang, Li, Lu, and Xie}]{zhang2024whisper}
Li~Zhang, Ning Jiang, Qing Wang, Yue Li, Quan Lu, and Lei Xie. 2024.
\newblock Whisper-sv: Adapting whisper for low-data-resource speaker verification.
\newblock \emph{Speech Communication}, 163:103103.

\bibitem[{Zuluaga-Gomez et~al.(2023)Zuluaga-Gomez, Prasad, Nigmatulina, Sarfjoo, Motlicek, Kleinert, Helmke, Ohneiser, and Zhan}]{zuluaga2023does}
Juan Zuluaga-Gomez, Amrutha Prasad, Iuliia Nigmatulina, Seyyed~Saeed Sarfjoo, Petr Motlicek, Matthias Kleinert, Hartmut Helmke, Oliver Ohneiser, and Qingran Zhan. 2023.
\newblock How does pre-trained wav2vec 2.0 perform on domain-shifted asr? an extensive benchmark on air traffic control communications.
\newblock In \emph{2022 IEEE Spoken Language Technology Workshop (SLT)}, pages 205--212. IEEE.

\bibitem[{Öpengin(2020)}]{opengin_2020_3744541}
Ergin Öpengin. 2020.
\newblock \href {https://doi.org/10.5281/zenodo.3744541} {\emph{Kurdish}}, chapter~21. Language Science Press.

\end{thebibliography}
\bibliographystyle{acl_natbib}

\appendix

% \section{Example Appendix}
% \label{sec:appendix}

% This is a section in the appendix.

\end{document}